\documentstyle[12pt,aaspp4]{article}
\makeatletter

\newlength{\bibhang}
\let\@internalcite\cite
\def\cite{\@ifstar{\citeyear}{\citefull}}
\def\cite{\let\@citeleft(\let\@citeright)%
    \@ifstar{\citeyear}{\citefull}}
\def\citenp{\let\@citeleft\relax\let\@citeright\relax
    \@ifstar{\citeyear}{\citefull}}
\def\citefull{\def\astroncite##1##2{##1~##2}\@internalcite}
\def\citeyear{\def\astroncite##1##2{##2}\@internalcite}

\def\@citex[#1]#2{\if@filesw\immediate\write\@auxout{\string\citation{#2}}\fi
  \def\@citea{}\@cite{\@for\@citeb:=#2\do
    {\@citea\def\@citea{; }\@ifundefined
       {b@\@citeb}{{\bf ?}\@warning
       {Citation `\@citeb' on page \thepage \space undefined}}%
{\csname b@\@citeb\endcsname}}}{#1}}

\def\@cite#1#2{\@citeleft#1\if@tempswa , #2\fi\@citeright}
\def\@biblabel#1{}

\makeatother

\newcommand{\PSbox}[3]{\mbox{\rule{0in}{#3}\includegraphics{#1}\hspace{#2}}}
\newcommand{\FigNum}[1]{\unitlength 1pt \begin{picture}(55,10)(-400,35) 
                        \put(0,0){Figure #1}
                        \end{picture}}
\newcommand{\msun}{$M_\odot$} 
\newcommand{\zsun}{$Z_\odot$} 
\newcommand{\persec}{\mbox{$\second^{-1}$}}
\newcommand{\percm}{\mbox{$\cm^{-2}$}}
\newcommand{\ppm}{\mbox{$\pm$}}
\newcommand{\cgsflux}{\erg\percm\persec}
\newcommand{\cgslum}{\erg\persec}

\newcommand{\approxlt}{\mbox{$\lesssim$}}
\newcommand{\approxgt}{\mbox{$\gtrsim$}}
\def\etal{{et~al.}}

\newcommand{\nh}{\mbox{$N_{\rm H}$}}
\newcommand{\nhtt}{\mbox{$N_{\rm H, 22}$}}
\newcommand{\ud}[2]{\mbox{$^{+ #1}_{- #2}$}}
\newcommand{\ee}[1]{\mbox{$10^{#1}$}}
\newcommand{\tee}[1]{\mbox{$\times 10^{#1}$}}

\newcommand{\perval}[2]{{#1\mbox{$^{#2}$}}} 

\def\x1608{{4U~1608$-$522}}
\def\nmon{{A0620-00}}
\def\cenx4{{Cen~X$-$4}}

\def\saxj1808{{SAX J1808.4$-$3658}}
\newcommand{\keV}{\mbox{$\rm\,keV$}}

\newcommand{\cm}{\mbox{$\rm\,cm$}}

\newcommand{\second}{\mbox{$\rm\,s$}}

\newcommand{\yr}{\mbox{$\rm\,yr$}}
\newcommand{\GramPerCc}{\mbox{$\rm\,g\,cm^{-3}$}}
\newcommand{\K}{\mbox{$\rm\,K$}}
\newcommand{\erg}{\mbox{$\rm\,erg$}}

\newcommand{\kteff}{$kT_{\rm eff}$}

\newcommand{\chandra}{{\em Chandra\/}}
\newcommand{\conx}{{\em Con-X}}
\newcommand{\rosat}{{\em ROSAT\/}}
\newcommand{\asca}{{\em ASCA\/}}

\newcommand{\beppo}{{\em BeppoSAX\/}}
\newcommand{\sax}{\beppo}

\newcommand{\Tcore}{$T_{\rm core}$}
\newcommand\plfx{\mbox{$F_{\rm X, PL}$}}

\newcommand\ktH{\mbox{$kT$}}
\newcommand\norm{\mbox{$N_\alpha$}}
\newcommand\chisqrnu{\mbox{$\chi^2_\nu$}}
\slugcomment{\it ApJ, Submitted Dec 5 2000; Accepted Dec 18 2000}

\begin{document}

\title{The Quiescent X-Ray Spectrum of the Neutron Star in \cenx4 
Observed with \chandra/ACIS-S} 

\author{Robert E. Rutledge\altaffilmark{1,6}, 
Lars Bildsten\altaffilmark{2}, Edward F. Brown\altaffilmark{3}, 
George G. Pavlov\altaffilmark{4,6}, 
\\ and Vyacheslav  E. Zavlin\altaffilmark{5,6}}
\altaffiltext{1}{
Space Radiation Laboratory, California Institute of Technology, MS 220-47, Pasadena, CA 91125;
rutledge@srl.caltech.edu}
\altaffiltext{2}{
Institute for Theoretical Physics and Department of Physics, Kohn Hall, University of 
California, Santa Barbara, CA 93106; bildsten@itp.ucsb.edu}
\altaffiltext{3}{
Enrico Fermi Institute, 
University of Chicago, 
5640 South Ellis Ave, Chicago, IL  60637; 
brown@flash.uchicago.edu}
\altaffiltext{4}{
The Pennsylvania State University, 525 Davey Lab, University Park, PA
16802; pavlov@astro.psu.edu}
\altaffiltext{5}{
Max-Planck-Institut f\"ur Extraterrestrische Physik, D-85740 Garching,
Germany; zavlin@xray.mpe.mpg.de}
\altaffiltext{6}{
Institute for Theoretical Physics, Kohn Hall, University of 
California, Santa Barbara, CA 93106}

\begin{abstract}

We report on spectral and intensity variability analysis from a
\chandra/ACIS-S observation of the transient, type-I X-ray bursting
low-mass X-ray binary Cen X-4. The quiescent X-ray spectrum during
this observation is statistically identical to one observed previously
with \sax, and close, but not identical, to one observed previously
with \asca. The X-ray spectrum is best described as a pure Hydrogen
atmosphere thermal spectrum plus a power-law component that dominates
the spectrum above 2 keV. The best-fit radius of the neutron star is
$r=$12.9\ppm2.6 $(d/1.2 \; {\rm kpc})$ km if the interstellar
absorption is fixed at the value implied by the optical reddening.
Allowing the interstellar absorption to be a free parameter yields
$r=$19\ud{45}{10} $(d/1.2 {\rm kpc})$ km (90\% confidence). The
thermal spectrum from the neutron star surface is inconsistent with a
solar metallicity. We find a 3$\sigma$ upper-limit of root-mean-square
variability $\leq 18\%$ (0.2-2.0 keV; 0.0001-1 Hz) during the
observation.  On the other hand, the 0.5-10.0 keV luminosity decreased
by 40\ppm8\% in the 4.9 years between the \asca\ and \chandra\
observations.  This variability can be attributed to the power-law
component. Moreover, we limit the variation in thermal temperature to
\approxlt10\% over these 4.9 years. The stability of the thermal
temperature and emission area radius supports the interpretation that
the quiescent thermal emission is due to the hot neutron star core.

\end{abstract}


\section{Introduction}

  Measuring both the mass ($M$) and the radius ($R$) of a neutron star
(NS) would strongly constrain the nuclear equation of state. To
distinguish between the competing models, several measurements of a
few percent accuracy in $M$ and $R$ are required.  While the masses of
several NSs have been constrained to 10\% or better (see
\citenp{thorsett99} for a recent review) from pulsar timing of Doppler
shifts, measuring the NS radii has proven to be difficult.  There are
at least five ways to measure $R$ using X-ray emission from accreting
NSs: (1) the spectral evolution of radius-expansion type-I X-ray
bursts; (2) the measurements of $M/R$ from the gravitational red-shift
of metallic spectral lines during type I X-ray bursts or $\gamma$-ray
lines during accretion; (3) inferring constraints from NS kHz
quasi-periodic oscillations; (4) pressure broadening and red-shift of
photospheric metal lines; and (5) spectral analysis of transiently
accreting NSs in quiescence.

Efforts to measure the NS radius through the spectral evolution of
radius-expansion type-I X-ray bursts (for a review, see \citenp{lvt93})
based on theoretical non-Planckian spectra
\cite{jvp82,london86,pavlov91,madej91,tit94} have been somewhat
successful. Performing this measurement with the emission from a
type-I X-ray burst has the advantage that the observed luminosity
originates from the NS photosphere and not the surrounding accretion
disk. The measured NS radii range from 6 to 15 km.  The reliability of
these measurements is limited by systematic uncertainties in the
emergent spectrum, the fraction of the NS surface area involved in the
burst and the elemental composition of the photosphere, as well as the
distance to the NS \cite{jvp87b,damen89,damen90}.

Attempts to measure NS photospheric metal lines as expected from model
atmospheres \cite{foster87} have had mixed
success. Absorption lines observed in the tails of type-I X-ray bursts
\cite{waki84,nakamura88,magnier89} have not been confirmed through
repeated observation with more sensitive instrumentation, resulting in
metallicity limits of $Z<0.4$\zsun\ \cite{day92}.  Moreover, the
interpretation of these lines as due to absorption in the photosphere
is in doubt \cite{madej90}.  Repeated observations of such lines, and
their identification with a known transition would measure
$M/R$. Gamma-ray lines from the accretion process are also a
possibility \cite{bildsten92,bildsten93}, though the levels of
emission are still below the ability of current instruments.

 Observations of kilo-Hz quasi-periodic oscillations from accreting
neutron stars, interpreted as the orbital period at or above the
marginally stable orbit \cite{kluzniak90,kaaret97,kluzniak98} have
been used to constrain both $M$ and $R$ using observations from eight
accreting neutron stars, finding $M\approx$2\msun\ \cite{zhang97b}. The
NS radius is then less than the marginally stable orbit for a mass of
this size ($<$18 km for a non-rotating star).

 Another method of measuring $R$ for a slowly rotating NS is the
approach described by Paerels \cite*{paerels97}, with metal lines
from an emergent NS X-ray spectrum. Measuring the pressure broadening
($\propto M/R^2$) and photospheric red-shift ($\propto M/R$) yields
both $M$ and $R$.  This method is intriguing because it is {\em
independent of distance}, which can be uncertain by up to a factor of
2 in the isolated transient NSs, although it can be determined to
$\approx$5\% for the low-luminosity X-ray sources in globular clusters
\cite{hertz83,verbunt84}. Once several metal lines are identified in
the photospheric spectrum, this method is very likely to provide a
number of $M$ and $R$ measurements.

 To these approaches has recently been added a new means of measuring
$R$ with X-ray spectroscopic observations of transiently accreting,
low magnetic-field ($<$\ee{10} G) NSs in quiescence
\cite{brown98,rutledge99,rutledge00}. Until this work, the
spectral observations had been interpreted as black-body radiation
\cite{jvp87,garcia94,verbunt95,asai96b,asai98,campana98a,garcia99,campana00},
yielding effective temperatures of $kT_{\infty}$=0.2-0.3 keV. For most
sources, the thermal component has a black-body emission area radius
much smaller than a NS ($\approxlt 1$ km).

 As described by Brown, Bildsten \& Rutledge \cite*[BBR98
hereafter]{brown98}, a black-body spectrum is not appropriate for
weakly magnetic\footnote{In low magnetic field NSs ($\le$\ee{10} G),
the magnetic field plays no role in determining the opacity of the
atmosphere at energies above 0.1 keV, and thus can be neglected
\cite{zavlin96}. } transiently accreting NSs with $kT_{\rm
eff}<5\tee{6} K$.  At accretion rates $\ll 10^{12} {\rm g \ s^{-1}}$
the accreting metals gravitationally settle faster than they are
supplied, and the atmosphere is nearly pure hydrogen
\cite{bildsten92}.  The dominant opacity is free-free absorption
($\propto \nu^{-3}$) which results in a spectrum in which higher
energy photons escape from deeper in the NS atmosphere, where the
temperature is higher \cite{pavlov78,rajagopal96,zavlin96}. This
hardens the spectrum which, (mis-)interpreted as a black-body, results
in systematically higher temperatures and {\em lower} emission area
radii. When the \asca\ and \rosat\ observations are fit with the more
appropriate thermal H atmosphere spectrum, the emission area radii are
$\approx$10 km \cite{rutledge99,rutledge00}, confirming that this
emission is mostly thermal. These objects have thus become the focus
of our efforts to measure the NS surface area.

If the emission area radius of the thermal part of the quiescent X-ray
spectrum is the NS radius, it should be constant from observation to
observation.  In addition, if accretion onto the NS makes no
contribution to the quiescent luminosity, the temperature should also
be constant on timescales $\ll$\ee{6} yrs, the core cooling timescale
(see \citenp{rutledge01} for exceptions).  In the most sensitive
observations, an additional power-law component is observed above a
few keV \cite{asai96b,asai98,campana98a,campana00}.  If produced
through interaction between the accretion and magnetosphere
\cite{campana98a}, it might vary on short ($\sim$msec) timescales, and
perhaps show a pulsation at the NS spin frequency.

 In this paper, we present X-ray spectroscopic analysis of \cenx4\ in
quiescence. \cenx4\ is a transient, type-I X-ray bursting low-mass
X-ray binary (for reviews on transients, see
\citenp{tanakalewin95,tanaka96,chen97}).  Its distance is estimated to
be 1.2\ppm0.3 kpc on the basis of two observed radius-expansion bursts
\cite{chev89}; formally, this is a distance upper-limit. It is in a
$P_{\rm orb}= 0.629$ day binary with a K dwarf
\cite{chev89,mcclintock90}, and has been observed in outburst twice
(1969 and 1979; see discussion in \S~\ref{sec:recur}).  We compare the
observed X-ray spectrum with a number of models, in particular with
the H atmosphere model. We also compare the spectrum with spectra of
the same source in archived observations. We place limits on changes
in the thermal spectrum of the NS on year long timescales. In
\S\ref{sec:anal}, we present the analysis of the \chandra\ data.  In
\S\ref{sec:multi}, we compare the \chandra\ data with past
observations, and characterize the spectral differences between this
and past X-ray observations.  We discuss these results and their
implications for this and other sources and conclude in
\S\ref{sec:discuss}.

\section{Observation and Analysis}
\label{sec:anal}

The observation occurred 23 Jun 2000 01:23:01-04:32:54 TT for a total
exposure time of 9561.4 seconds with \chandra/ACIS-S3 (backside
illuminated). The source position was aimed 4\arcmin\ off-axis, with a
1/8 subarray used with 0.4 sec exposures. The time between successive
frames was 0.44104 sec, which gives $\approx$ 10\% dead-time.  We
analyzed data from the pre-processed Level 2 FITS data provided with
the standard data products.  The X-ray source appears at the known
optical position of Cen X-4 (\ppm1 \arcsec; \citenp{canizares80} ).
Only one other point source (previously unknown) was found in the
field with the \chandra\ Interactive Analysis of Observations Software
(CIAO) point-source detection tool {\em celldetect}~\footnote{available at http://asc.harvard.edu}, with a S/N of 3.5,
roughly 1\arcmin\ away.

We extracted data from a circle 10 pixels in radius about the \cenx4\
X-ray source position, and background from an annulus with inner and
outer radius of 13 and 50 pixels, respectively.  There were a total of
2714 good counts in the source region, and 575 in the background
region; we expect 25 background counts in the source region
($\approx$1\% of the total counts in the source region).

\subsection{\chandra\ Spectral Analysis}

The temperature of the focal-plane instruments has been decreased
during \chandra's lifetime; this alters the energy response of the
ACIS-S chips. The data were taken with a focal-plane chip temperature
of $-120{\rm C}$, and we used the corresponding response files for
this temperature, according to the standard ACIS-S analysis.

  We binned the photons between 0.5 and 1.5 keV into bins of width
$\approx$130 eV wide, comparable to the energy resolution (FWHM) in
the BI-S3 detector. Above 1.5 keV, we binned the data so that there
were 40 counts per bin (wider than the spectral resolution).  We used
spectral data in the energy range 0.5-10 keV.  While there are
significant counts below 0.5 keV, the ACIS-S energy calibration is
presently not reliable below this energy; we will re-examine this
analysis when the energy calibration is refined below 0.5 keV.  In
addition, the current \chandra\ response files for the ACIS-S-BI
underestimate the detector area by up to 20\% near 0.5 keV (N. Schulz,
priv. comm.). To account for the calibration uncertainties, we
included a 25\% systematic uncertainty in the 0.5-0.6 keV energy
range, and a 5\% systematic uncertainty in 0.6-0.7 keV energy
range. This is in addition to a 4\% systematic uncertainty across all
energies that accounts for other calibration uncertainties.

  We fit the data in XSPEC v11 \cite{xspec}, with several spectral
models (powerlaw, H atmosphere, blackbody, Raymond-Smith, or thermal
bremsstrahlung), all with galactic absorption (\nh) as a free
parameter, except where noted.  The H atmosphere spectrum is not a
standard XSPEC model, but has been described and calculated elsewhere
\cite{zavlin96}.  The hydrogen column density for \cenx4\ was
estimated during quiescence to be \nhtt$<$0.20 (\citenp{asai96a};
\nh=\ee{22}\nhtt\ \perval{cm}{-2}); its optical reddening is
$E(B-V)=0.1$ \cite{blair84}.  These values are consistent with an
approximate optical reddening/hydrogen column density ratio found from
observing the halos of X-ray sources \cite{gorenstein75,predehl95},
which imply an equivalent hydrogen column density of \nhtt=0.055.

No single component model fit the data acceptably (prob $\ll
10^{-6}$). In the black-body and H atmosphere fits, the high-energy
($>$2 keV) powerlaw spectral component reported previously
\cite{asai96b,campana00} is apparent. In general, the single component
models failed due to the presence of this high-energy component.  We
then fit the data with two-component models, where the second
component is a power-law that accounts for the $>$2 keV emission.  The
best-fit absorbed H atmosphere spectrum has parameters listed in
Table~\ref{tab:chandra}.  We also provide the best fit parameters with
\nh\ held fixed at the optically implied value (\nhtt=0.055), and the
resulting intrinsic (that is, unabsorbed) spectrum in
Figure~\ref{fig:chandraspec}.

While a double power-law model was statistically acceptable, the low
energy component has a steep power-law slope (photon index
$\alpha$=6.2\ud{1.7}{0.9}, 90\% confidence; all uncertainties and
upper-limits are 90\% confidence unless otherwise stated) and
considerably higher absorption than observed previously from this
source or implied by optical observations (\nhtt=0.50\ud{0.22}{0.10}).
We reject the model on this basis.

  A Raymond-Smith plasma model \cite{raymondsmith} with an underlying
power law is also statistically acceptable, however, with an abundance
limited to $Z<$1.4\tee{-3}$Z_{\rm sol}$. \footnote{While the
metallicity of the companion has not been observationally constrained,
Martin \etal \cite*{martin94a}, while investigating Li abundance,
found that the Ca I $\lambda6717$ line is consistent with solar
metallicity, and that an assumed  solar Fe abundance in their
spectral fits to the Li I doublet is consistent with the data.  In
optical spectral studies of the companion
\cite{cowley88,chev89,mcclintock90,mcclintock00}, while metal
absorption features are observed, the metallicity of the companion is
not estimated; however, it does not appear to be so substantially
sub-solar.}  In addition, it has been shown previously that the
X-ray-to-optical flux ratio is substantially greater than is typical
for stellar coronal sources \cite{bildsten00}.  We therefore reject a
model which ascribes this emission to the corona of the stellar
companion.  Though not physically motivated, a thermal bremsstrahlung
spectrum acceptably fits the data, with a volume emission measure of
$\int n_e\; n_I dV = 2.3\tee{56} (d/1.2 {\rm kpc})$ \perval{cm}{-3}
and a temperature of 0.34 keV.

To enable comparison with previous work, we also fit the data with an
absorbed blackbody and powerlaw.  The best fit with a fixed \nhtt=0.26
\cite{campana00} is rejected (\chisqrnu=3.1/10 dof; prob=5\tee{-4}).
With all parameters free, we find \nhtt$<$0.071 (90\%);
\kteff=0.175\ud{0.012}{0.017}; r=1.1\ud{0.4}{0.2} $(d/1.2 \;  {\rm
kpc})$ km; $\alpha=$1.2\ud{0.4}{0.5} (\chisqrnu=0.75/9 dof; prob=0.67).

\subsection{Solar Metallicity Atmosphere}

We fit the \chandra\ data with a solar metallicity atmospheric model
for a 10 km, 1.4\msun\ NS, with relative abundances given by Grevesse
\& Noels \cite*{grevesse93}. A galactic absorption and solar
metallicity atmosphere model failed to fit the spectrum
(\chisqrnu=18/11 dof), due to the high energy ($>$2 keV) excess
attributed above to the power-law component, and to spectral curvature
below 1 keV. With an additional power-law component, the best-fit
spectrum is still unacceptable (prob = 3\tee{-8}), largely due to the
Fe L edge near 0.534 keV (for a 1.4\msun, 10 km NS).  While the model
atmospheres assume no rotation and are static, the major discrepancy
between the model and data is this absorption edge, which will not be
significantly altered by these effects.

A solar metallicity NS atmosphere spectrum is therefore strongly
rejected. More detailed investigation regarding sub-solar metallicity
spectra is in progress, and will be presented in forthcoming work.  It
would be highly useful to have response matrices refined down to 0.3
keV for this work. These will be valuable limits, as accretion at a
rate high enough to explain the quiescent thermal emission would
enrich the atmospheric metal content to a detectable level (BBR98). 

\subsection{Comparison between the \chandra,  \sax\ and \asca\ X-ray spectra}
\label{sec:multi}

  \cenx4\ has been detected on 6 occasions in quiescence,
with {\em Einstein} and {\em EXOSAT} \cite{jvp87}, twice with \asca\
\cite{asai96b,asai98}, once with \rosat/HRI \cite{campana97} and once
with BeppoSAX \cite{campana00}.  The measured luminosities and
blackbody spectral parameters from the published observations are
listed in Table~\ref{tab:prevwork}.

  In this section, we compare the \chandra\ X-ray spectrum with that
previously obtained with \sax\ \cite{campana00} and with the \asca/SIS+GIS
spectrum \cite{asai96a,asai98,rutledge99}.  We do not compare with one
other existing \asca/GIS observation as it was performed largely with
the GIS, with lower S/N than the first \asca\ observation, and is
comparable to the first \asca\ observation \cite{asai98}.  We have
previously re-analyzed the \asca\ observation \cite{rutledge99} and
use the same resulting spectrum here.

\subsubsection{Re-Analysis of \sax\ observation}
\label{sec:sax}

 Campana \etal (2000; C00 hereafter) interpreted the quiescent X-ray
spectrum (measured with \asca\ and \sax) of \cenx4\ as a black-body,
and found a mean blackbody radius 3.1\ud{3.7}{3.1} km for the thermal
component (the best-fit blackbody radius with \sax\ data alone was
r=10\ud{10.8}{1.8} km).  However, \nh\ was held fixed at the best-fit
value of \nhtt=0.26, which is higher than the optically implied value.
We undertook a reanalysis of the \sax\ observation of \cenx4,
previously analyzed by C00. There are two major differences between
our spectral analysis of the LECS data (0.3-2 keV) and theirs:

\begin{enumerate}

\item For the LECS background data, C00 used the standard ``blank
fields'' background which are taken from several ``blank'' areas of
the sky \cite{parmar99}. However, \cenx4\ is located in a region of
the sky with higher than average low-energy background
($l=332.24\deg$, $b=+23.88\deg$), according to the
ROSAT/All-Sky-Survey Soft X-ray Background maps ($\approx$0.75 keV;
\citenp{snowden97}, Fig 6b).  Thus, the ``blank fields'' background
underestimates soft X-ray background in the \cenx4\ observation.  We
used an alternative method of background estimation described by
Parmar \etal\ ( the ``annulus'' method; \citenp*{parmar99}), in which
the background intensity is assumed to be proportional to the
countrate in an annulus in the LECS detector, with a spectrum
identical to that found with the ``blank fields'' method.  By this
method, the 0.3-1.0 keV background counts account for 60\% of the
total counts in the \cenx4\ source region, whereas they only account
for 34\% in the ``blank fields'' method; the counting uncertainty in
the source region is 10\%, and therefore this difference in background
countrate is significant.  As the ``annulus'' method scales with the
local background countrate, it is a more realistic estimation of the
local low-energy ($<$2 keV) background, and so we adopt it.

\item For the LECS data, we used a larger extraction radius than C00
(8 \arcmin\ vs. 4\arcmin); the LECS point-spread-function becomes
broad toward the lower energies (80\% encircled energy at 1.5 keV is
3\arcmin .0, and at 0.28 keV is 6\arcmin .1; the 95\% radius is
5\arcmin .5 and 8\arcmin .5 at 1.5 and 0.28 keV, respectively).  The
smaller extraction radius used by C00 may have biased the resulting
\nh\ upwards.  This collected 647 counts from the LECS in 21524 sec
(much greater than the 233 counts found by C00). In the 0.3-1.0 keV,
there are 84 counts in the source region. We estimate that using an
8\arcmin\ radius instead of a 4\arcmin\ radius increases the observed
source countrate by 30\%, which is larger than the Poisson uncertainty
of 10\%.

\end{enumerate}

For the MECS data, we used a 4\arcmin\ extraction radius, which
extracted 651 counts (comparable to the 632 found by C00).  We used
the standard MECS background file (MECS23\_bkg.evt) extracting
background counts using the same region as for the source fields.  The
source background subtracted countrates are (5.3\ppm1.6)\tee{-3} c/s
(0.1-3.1 keV) and (3.3\ppm0.4)\tee{-3} c/s (1.7-9.0 keV) in the LECS and MECS
instruments, respectively.  The LECS data has lower S/N than reported
by C00 due to the higher background;  we find a slightly higher
MECS countrate than found previously.

For comparison with previous work, we fit the \sax\ LECS+MECS data
with an absorbed black-body plus power-law spectrum.  We obtained an
acceptable fit (\chisqrnu=0.65/1 dof; prob=0.42), with the following
best-fit parameters: \nhtt$<$0.95, $\alpha$=1.6\ud{0.6}{0.8},
\kteff=0.10\ud{0.15}{0.04} keV.  We were unable to put any reasonable
limits on the BB radius ($>$0.4 km, unbounded from above), as the best-fit
spectrum merely would increase the column density to compensate for
larger and larger areas.  When we hold the column density fixed at the
best-fit value, we find $r_{BB}$=7\ud{26.3}{6.5} km; when we hold the
column density fixed at the optically implied value (\nhtt=0.055), we
find $r_{BB}$=2\ud{5}{1.6} km.  These values are comparable to those
found with \asca\ \cite{asai96b,rutledge99}, and below those found by
C00.  We attribute this difference to our more accurate background
subtraction and larger extraction radius for the LECS data.  The
unabsorbed total flux is 2.7\tee{-12} \cgsflux (4.7\tee{32} \cgslum ,
0.5-10.0 keV); the unabsorbed flux of the BB component is 2.2\tee{-12}
\cgsflux\ (0.5-10.0 keV), and of the power-law component it is
0.44\tee{-12} \cgsflux\ (0.5-10.0 keV).  As we find in the next
section, the \sax\ and \chandra\ spectra (and their corresponding
fluxes) are, within statistics, the same.

\subsubsection{Joint Spectral Fitting of \chandra,  \sax\ and \asca\ X-ray Spectra}

We performed a joint spectral fit of the \chandra, \asca, and \sax\
data.  We assumed a 4\% systematic uncertainty in the detector
responses for all instruments, and held \nh\ fixed at its optically
implied value. We used an absorbed H atmosphere+powerlaw spectrum.
The best spectral fit rejects a single model to account for all three
observed spectra (\chisqrnu=3.3/53 dof; prob=5\tee{-15}).  A joint
\chandra/\sax\ fit provided a statistically acceptable fit, indicating
that the \chandra\ and \sax\ spectra are statistically identical.  A
joint \chandra/\asca\ fit was not acceptable.  The \asca\ data are, by
themselves, acceptably fit by this same assumed spectrum
\cite{rutledge99}.

We jointly fit the \chandra\ and \asca\ spectra, permitting one of the
five parameters (\ktH, $r$, $\alpha$, and power-law normalization
\norm) to be different between the two spectra, each in turn.  None of
these fits are formally statistically acceptable.  However, while a
change in $\alpha$ is soundly rejected (prob$\leq$4\tee{-10}), and a
change in $r$ is of low probability (prob =0.001), changes in the
\norm\ or \ktH\ are found to be marginal (both prob = 0.01).  This is
in the range where systematic uncertainty in the detector responses
(in particular, the area as a function of energy) become important.
For example, if we increase the systematic uncertainty from 4\% to 8\%
(which, for example, could be due to a systematic offset in the
absolute flux calibrations between the two detectors), the best fits
for a changing \norm\ or \ktH\ become statistically acceptable.

Thus, while the \chandra\ and \asca\ spectra are significantly
different, we cannot unequivocally state whether this difference is
due to a change in the \norm\ or the \ktH , a combination of these
with other parameters, or a systematic difference between the absolute
calibrations of the two instruments. However, we provide the best-fit
spectral parameters in Table~\ref{tab:jointfit}, for a changing power-law
normalization and thermal temperature.  These then serve as an
upper-limit for variability in these parameters between the two
observations.  Between the \chandra\ and \asca\ epochs, then, we find
that the thermal temperature decreased (at most) from \ktH\ =0.085 to
0.077 keV; or the power-law normalization decreased from 15.2\tee{-5}
to 5.9\tee{-5} phot \perval{cm}{-2} \perval{s}{-1} at 1 keV.
Moreover, the interpretation of these spectra impacts the best-fit
power-law slope, which is $\alpha$=1.0 if it is the temperature which
changes, or $\alpha$=1.7 if it is the normalization which changes.

Finally, we fit a spectral model to the \chandra, \sax, and \asca\
data assuming that the power-law slope and flux changes, but that the
thermal component does not.  The resulting best-fit parameters and
uncertainties are shown in Table~\ref{tab:3fit}.  The best-fit is
statistically acceptable.  The power-law component (taken to the be
the same for the \chandra\ and \sax\ observations) is flatter in the
\chandra+\sax observations than during the \asca\ observation
(1.0\ppm0.4 vs. 1.9\ppm0.3), and has a lower flux, by about a factor
of two.

We also fit these spectra assuming that the thermal $r$ and \ktH\ vary
while the power-law component does not (although this is not
theoretically motivated).  An acceptable fit is found
(\chisqrnu=1.33/52 dof; prob=0.06), with \chandra+\sax values of
($r$=13.4\ud{4.2}{2.4} km , \ktH=0.074\ud{0.006}{0.008} keV) and
\asca\ values of ($r$=7.1\ud{2.5}{3.0} km,
\ktH=0.10\ud{0.026}{0.009}keV).

\subsection{Intensity Variability}

Variability in the luminosity of \cenx4\ has been observed over
timescales from days to years.  Van Paradijs \etal\ \cite*[JVP87
hereafter]{jvp87} detected \cenx4\ using {\em EXOSAT}/LE1 with the CMA
instrument and LEXAN 3000 \AA\ filter \cite{taylor81}, with a countrate
of (5.4\ppm1.2)\tee{-3} c/s.  JVP87 assumed a thermal bremsstrahlung
spectrum with temperature between 1 and 5 keV and a column density
\nhtt=0.066, which, for this countrate, corresponded to a flux at
earth of (2.2-5.6)\tee{-12} ${\rm erg \ s^{-1} \ cm^{-2}}$ . Using the
same assumptions with an earlier {\em Einstein}/IPC observation, they
found a flux at earth between (1.1-1.5)\tee{-12} ${\rm erg \ s^{-1} \
cm^{-2}}$, concluding that the X-ray luminosity must have increased
between the Einstein and {\em EXOSAT } observations by 50-500\%.
Campana \etal\ \cite*{campana97} measured a flux variation of a factor
of 3 in $<$4 days using \rosat/HRI.

 We started by looking for variability during our observation.  We
first separated the data into two energy bins: pulse-invariant (PI)
bins 2-136 (0.2-2 keV) and PI bins $>$136 ($>$2 keV).  We included PI
bins in which the area and energy calibration is presently not
reliable; while these are not useful (presently) for flux and spectral
calculations, they can be used for investigations in variability.

There were a total of $N_{\rm phot}=2553$ counts in PI bins 2-136
($\approx$0.2-2 keV).  We used 23903 time bins which were 0.44104 sec
in length which each had exposure of only 0.40 sec, for a total
observation time of $T_{\rm obs}=10542$ sec.  We performed a Fourier
transform, producing a power density spectrum (PDS) of 11951 frequency
bins across the 0.0001-1.13 Hz frequency range, normalized according
to Leahy \etal\ \cite*{leahy83}.  We logarithmically rebinned this
data, and fit the PDS with functions $P(\nu)$, to extract the
integrated r.m.s power.  The PDS is acceptably described as counting
noise (constant power of 2.0; \chisqrnu=1.41/16 dof; prob=0.13); we
therefore observe no variability in this data.  Using a power-law
distribution with the slope of $\alpha=1$ held fixed, with an
underlying Poisson level ($P=2$) also held fixed, we find a 3$\sigma$
upper limit on the root-mean-square variability of $<$18\% (0.0001-1
Hz).  For a flat power-law ($\alpha=0$), the 3$\sigma$ upper limit is
$<$10\%.

There are a total of 151 counts in PI bins $>$136 (\approxgt 2.0 keV)
in this observation.  We performed an identical PDS analysis as for
the low energy counts, although in the fit to the PDS we held the
exponent fixed at $\alpha$=1.0 (producing an acceptable fit;
\chisqrnu=1.46/15 dof).  The 3$\sigma$ upper-limit to the RMS variability
$>$2.0 keV is $<$50\% (0.0001-1 Hz).

We also looked for longer term variability by 
investigating the hypothesis that the \chandra\ and \asca\ spectra
were identical in all parameters, but different in absolute
normalization.  We found that the spectra were describable in this way
(\chisqrnu=1.05/47 dof; prob=0.37), with the \asca\ spectrum a factor
1.65\ppm0.12 (90\% uncertainty) more luminous than the \chandra\
spectrum.  We interpret this as fitting a well-constrained spectrum
from \chandra\ (signal-to-noise ratio S/N$\approx$100) to the low S/N
data of \asca\ (S/N$\approx$4).  This provides a rough measure of the
luminosity difference between the \asca\ and \chandra\ observations --
a decrease of 40\ppm8\% over 4.9 yr. 

\section{Summary, Conclusions and Implications}
\label{sec:discuss}
\label{sec:con}

  We have analyzed the X-ray spectrum and intensity variability of
\cenx4\ observed with \chandra, and compared it with earlier
observations by \asca\ and \sax. The \chandra\ X-ray spectrum is
inconsistent with all single component models we applied, requiring at
least a two component model, with a power-law component that dominates
the spectrum above 2 keV and a softer (most likely thermal) component
which dominates the spectrum below this energy. The intensity
variability during the \chandra\ observation is $<$18\% (3$\sigma$) on
timescales between 1-10,000 sec in the 0.2-2.0 keV energy band
dominated by the thermal component. The limit on variability is weaker
above 2 keV ($<$50\% r.m.s., 3$\sigma$).

 Cen X-4's luminosity decreased between the \asca\ and \chandra\
observations (4.9 yrs) by 40\ppm8\% (0.5-10.0 keV). We attribute this
variation to changes in the power-law component. This is comparable to
the factor of $\sim$3 decrease found over a few days by Campana \etal\
\cite*{campana97}.  We cannot statistically exclude that the power-law
component remains constant, and the thermal component varies.  Thus,
\cenx4\ is variable in quiescence on timescales longer than hours.

\subsection{The Thermal Component} 

Rutledge \etal\ \cite*{rutledge99,rutledge00} have shown that the soft
emission seen in all NS transients in quiescence is best explained as
thermal emission from a pure hydrogen NS atmosphere. We have found the
same in this \chandra\ observation. Indeed, comparing X-ray spectra
taken with \chandra, \asca\ and \sax\ at different times over 4.9
years, we find that this thermal component is consistent with being
constant, with a best-fit radius of $r$=12.9\ppm2.6 $(d/1.2 {\rm
kpc})$ km, and a temperature \ktH=0.076\ppm0.007 keV. These results
improve the precision of the quiescent spectrum by a factor of $\sim$3
over previous work \cite{rutledge99}. We place an upper-limit on the
amount of temperature variability across 4.9 years of $\approxlt$10\%
(Table~\ref{tab:jointfit}). The agreement between this radius and that
measured during Type I X-ray bursts from other systems is strong
confirmation that the emitting area is the NS surface.

Two energy sources for the thermal emission have been discussed:
accretion in quiescence at a low-rate \cite{jvp87,menou99} and
re-emission of heat deposited into the crust during the large
accretion events (BBR98). The required accretion rate for the
bolometric luminosity of $7\times 10^{32} \ {\rm erg \ s^{-1}}$ is
$\dot M_q\approx 6\times 10^{-14} M_\odot \; {\rm yr^{-1}}$, adequate to
keep the metal content in the atmosphere comparable or larger than
that in the accreting material \cite{bildsten92}.  If the metal
content in the accreting material is very sub-solar (that is,
effectively pure hydrogen), accretion at these rates would give
thermal emission much like we observe \cite{zampieri95}. However,
solar metallicity accretion is likely ruled out, largely due to the
lack of absorption from the Fe L edge near 0.5 keV.  Tighter
constraints on the photospheric metallicity will be the subject of
future work with these data and will allow us to more thoroughly
constrain the active accretion hypothesis, which does not have a
specific way of predicting $\dot M_q$.

  The other possible mechanism is re-emission of heat deposited in the
inner crust due to pycnonuclear reactions, electron captures and
neutron emissions \cite{haensel90} during accretion events
(BBR98). The layers where this heat is deposited are in close thermal
contact with the NS core.  The reactions then heat the core to a
temperature of $\sim$\ee{8} K over $\sim$\ee{4} yr (BBR98;
\citenp{colpi00b}). The NS reaches an equilibrium where the time
averaged nuclear heating equals the quiescent thermal emission,
$L_q\approx (1 \ {\rm MeV}/{\rm nucleon}) \langle \dot M \rangle$,
implying a time averaged accretion rate for this NS of $\langle \dot
M\rangle\approx 10^{-11}M_\odot \; { \rm yr^{-1}}$. 

\label{sec:recur}
We can compare this $\langle \dot M\rangle$ to that implied by the
outbursts.  The outburst in 1969 had a total fluence of $\approx 3 \;
{\rm ergs \ cm^{-2}}$ \cite{chen97}. There has not been another large
outburst of this magnitude recorded since; the small outburst in 1979
had a fluence two orders of magnitude smaller. It is uncertain to
measure the time-averaged accretion rate on the basis of a single
outburst.  If a large outburst like that in 1969 occurs every 100
years, the time-averaged accretion flux would be $\approx 10^{-9} \;
{\rm ergs \ cm^{-2} \ s^{-1}}$, which at the 1.2 kpc distance gives
$\langle \dot M \rangle \approx 10^{-11} M_\odot {\rm yr^{-1}}$,
comparable to that implied by the level of the quiescent thermal
emission.  However, this is a low accretion rate for a binary at this
long orbital period, where the likely driver of mass transfer is
nuclear evolution of the binary (see \citenp{webbink83}). So, this is
an outstanding puzzle for this source. If outbursts like that seen in
1969 occur more frequently than 100 years, the time averaged accretion
rate would be higher and in conflict with that implied by the
quiescent thermal emission through the mechanism of BBR98, unless the
fraction of deposited energy re-emitted as photons is much less than
unity. Possible solutions to this, involving enhanced neutrino
emission from the core, have been proposed by Colpi \etal\
\cite*{colpi00b} and Ushomirsky \& Rutledge \cite*{rutledge01}.

If the thermal emission is due to a hot neutron star core, then the
measured effective temperature of $0.076\pm 0.007\keV$ tells us the
internal NS temperature. For example, if the outer layer consists of
light elements to a density $6\times10^{5}\GramPerCc$, then using the
fit of Potekhin \etal\ \cite*{potekhin97}, we find \Tcore$ =
(5.1\ud{1.2}{1.1})\tee{7}\K$.  For these core temperatures, the
modified URCA neutrino luminosity is orders of magnitude less than the
photon luminosity.  The NS core would not be in a thermal steady-state
if pion condensation occurs, as the resultant neutrino luminosity
\cite{umeda94} would be much larger than the heating supplied by
reactions in the deep crust. As a result, if some enhanced cooling
mechanism were operating in the core, then we would expect
\Tcore$\lesssim 10^7\K$ and the crust and surface temperature should
be decreasing on a timescale of $\sim 1\yr$ (BBR98;
\citenp{rutledge01}).

The lack of luminosity variability in general, and the stability of
the thermal component over a $\sim$5 year timescale supports the model
proposed by BBR98 for the thermal component. It remains to be
unequivocally demonstrated, however, that the variability observed is
restricted to the power-law component, and that the variability of the
temperature of the thermal component is limited to \approxlt 1\%, as
expected from variations in the \Tcore\ from outburst to outburst
\cite{colpi00b}. 

\subsection{The Power-Law Spectral Component}

The origin of the hard power-law component remains unclear.  Many
ideas have been put forward, ranging from emission due to an active
pulsar wind colliding with the accretion disk, x-ray emission from a
turned-on radio pulsar \cite{stella94} and accretion onto the
magnetosphere (for a review, see \citenp{campana98b}).

There is no direct knowledge of either the NS spin or magnetic field
in Cen~X-4, though the presence of Type I bursts points to $B<10^{10}
G$. None of the above models make specific predictions about either
the level of emission or its spectral shape which explain the present
observations.

 Moreover, the limitation to $<$1\% rms of coherent intensity
pulsations in the transiently accreting NS Aql~X-1 (another transient,
type-I X-ray bursting source) following a rapid decrease  in
flux \cite{chandler00} which was interpreted as due to magnetic
inhibition of the accretion flow \cite{campana98a,zhang98a} does not
support the idea that accretion onto the compact object is ultimately
restricted by a ``propeller effect'', favoring instead an
interpretation of the end of transient outbursts as due to an end of
the disk instability, such as occurs in dwarf novae \cite{king98}.  In
fact, abrupt declines in dwarf novae outburst fluxes are observed, and
are explained within the disk instability model \cite{cannizzo94},
without invoking magnetic inhibition.  The absence, then, of a
magnetosphere in transiently accreting NSs would preclude attributing
the power-law to magnetospheric accretion, and energy production would
have to be attributed to another site within the NS binary system.

Spectral comparison between the \asca\ and \chandra\ observation
indicate that a change in (only) the power-law normalization is
marginally acceptable, and therefore that a change in the spectral
slope (from $\alpha=1.9\ppm0.3$ to $\alpha=1.0\ppm0.4$) is also
required.  It is unclear how a change in the spectral slope could be
produced while limiting the luminosity decrease to less than a factor
of three in any of the proposed emission mechanisms \cite{campana98b}.

There might be some help in understanding the power-law component from
other wavelength bands, where it was found recently that, in
quiescence, the energy density ($\nu F_\nu $) is nearly flat from UV
through the X-ray energy range, in marked contrast to \nmon, in which
the energy density falls in the UV band \cite{mcclintock00}.  This
flat-spectrum has been interpreted as due to a shock at the
splash-point of accretion from the companion \cite{menou01}.  However,
while we also find a flat spectrum for the \asca\ observation (photon
power-law slope of $\alpha=1.9\ppm0.4$) we  find an {\em
increasing} energy density during the \chandra\ observation
($\alpha=1.0\ppm0.4$).  Clearly, further observational study of the
power-law component is required to understand even its most basic
aspects.

\subsection{Future Work on Neutron Star Radii from Thermal Emission} 

  The dominant systematic uncertainty in the measurement of the radius
of the neutron star in \cenx4\ is the distance to the object. This
will be dramatically improved by the Space Interferometric Mission,
which can measure the parallax of the 18$.^m$5 counterpart to
$\sim10\; \mu$asec, or $<$1\% distance uncertainty at 1.2 kpc.  This
would make \nh\ the dominant uncertainty in the measured thermal
emission radius, followed by any systematic uncertainty in the
modelled spectrum.  For example, a planned 50 ksec XMM observation
will permit constraint of $r/D$ to $\sim$2\% while simultaneously
measuring \nh.

It has been suggested that some fraction of the low-luminosity X-ray
sources in globular clusters are transiently accreting NSs in
quiescence \cite{verbunt84,rutledge00}. These make excellent targets
for NS radius measurements, as there are multiple objects per
observing field, all at the same distance and interstellar column
density (BBR98). \chandra\ imaging observations of globular clusters
\cite{grindlay00,pooley00} indicate source densities in excess of
several per square arcmin (down to luminosities of $L_X\sim$\ee{30}
\cgslum) with over 100 X-ray sources in the core of 47 Tuc alone.  It
is still unanswered as to what fraction of these are quiescent NS's.
The high source density will complicate X-ray spectroscopy with XMM or
\conx\ (with angular resolution $\sim$15\arcsec), making \chandra\
imaging spectroscopy the best way to pursue this science.

\acknowledgements

The authors are grateful to the \chandra\ Observatory team for
producing this exquisite observatory. R.R. thanks Andy Fabian for a
useful conversation regarding historical measurements of photospheric
absorption lines. The authors thank Dany Page and Andrew Cumming for
comments on the text prior to submission.  This research was partially
supported by the National Science Foundation under Grant
No. PHY99-07949 and by NASA through grant NAG 5-8658, NAG 5-7017 and
the \chandra\ Guest Observer program through grant NAS 8-39073.
L. B. is a Cottrell Scholar of the Research Corporation.
E. F. B. acknowledges support from an Enrico Fermi Fellowship.


%
\clearpage
\pagestyle{empty}
\begin{figure}[htb]
\caption{ \label{fig:chandraspec} The $\nu F_\nu$ model spectrum of
\cenx4, and the observed \chandra/ACIS-S BI data above 0.5 keV.  The
solid line is the best-fit {\em unabsorbed} H-atmosphere plus
power-law model spectrum with \nhtt=0.055 held fixed (that is, the
intrinsic X-ray spectrum of \cenx4, prior to absorption by the
inter-stellar medium; see Table~\ref{tab:chandra}).  The dashed line
is the model power-law component, and the dashed-dotted line is the
H-atmosphere component.  The two spectral components are equal near
$\approx$2~keV, above which the power-law component dominates, and
below which the H atmosphere component dominates.  The H atmosphere
spectral parameters are $r=$13.3\ud{2.4}{3.9} km, and
$kT$=0.074\ud{0.012}{0.005} keV.  The power-law component is rising in
$\nu F_\nu$.  The crosses are the observed \chandra\ data, with
error-bars in countrate.  }
\end{figure}

\clearpage
\pagestyle{empty}
\begin{figure}[htb]
\PSbox{fig1.ps hoffset=-80 voffset=-80}{14.7cm}{21.5cm}
\FigNum{\ref{fig:chandraspec}}
\end{figure}

\begin{deluxetable}{lc}
\scriptsize

\tablecaption{\label{tab:chandra} \chandra\ Observation Spectral Parameters (0.5-10 keV)}
\tablewidth{7cm}
\tablehead{
\colhead{} & 
\colhead{} \\
}
\startdata
\cutinhead{H Atmosphere + Power-Law}\\
\nhtt		& 0.10\ud{0.10}{0.07}	\\
$\alpha$	& 1.0\ppm0.4 \\
\plfx		&  5 	\\
$kT$ (keV) 	& 0.067\ppm0.019 \\
$r$ 		& 19\ud{45}{10}	\\
Total Model Flux &  13		\\
\chisqrnu/dof (prob) & 0.94/9 (0.49)	\\

\cutinhead{H Atmosphere + Power-Law (\nh\ fixed)}\\
\nhtt		& (0.055)	\\
$\alpha$	& 1.0\ud{0.6}{0.4}		\\
\plfx		&  4.6			\\
$kT$ (keV) 	& 0.074\ud{0.012}{0.005}	\\
$r$ 		& 13.3\ud{2.4}{3.9} km 	\\
Total Model Flux &  11.3			\\
\chisqrnu/dof (prob) & 0.83/10 (0.50)	\\

\enddata
\tablecomments{X-ray fluxes are un-absorbed, in units of \ee{-13}
\cgsflux\ (0.5-10 keV).  Assumed source distance d=1.2 kpc.}
\end{deluxetable}

\begin{deluxetable}{llcccc}
\scriptsize

\tablecaption{\label{tab:prevwork} \cenx4\ Quiescent Observations}
\tablewidth{17cm}
\tablehead{
\colhead{Ref.} & 
\colhead{Obs Date} &
\colhead{$kT_{BB}$} &
\colhead{X-ray Luminosity} &
\colhead{$\alpha$} &
\colhead{\nhtt} \\
\colhead{}	&
\multicolumn{1}{r}{Instrument} & 
\colhead{}	&
\colhead{(passband)}	&
\colhead{}	&
\colhead{}	\\
}
\startdata

1	& 
1980 Jul 28& 
0.32\ud{0.12}{0.08}&
2-3\tee{32}	&
\nodata 	& 
0.066	\\
		&
\multicolumn{1}{r}{{\em  Einstein}/IPC } & 
		&
 (0.5-4.5 keV) 	&	
		&
		\\

1	& 
1986 Feb 21	&
\nodata		&   
4-11\tee{32}	&
\nodata		& 
0.066 \\
		&
\multicolumn{1}{r}{{\em EXOSAT}/CMA 	}&
		&
 (0.5-4.5  keV) 	&	
		&
		\\

2 	& 
1994 Feb 27-28	&
0.16\ud{0.03}{0.02}	& 
4\ud{4}{1.7}\tee{32}	&
1.9\ppm0.3	&
$<$0.2 \\
		&
\multicolumn{1}{r}{\asca/SIS+GIS	}& 
		&
 (0.5-10  keV) 	&	
		&
		\\

3		&
1995 Aug 16-26	&
\nodata		&
7\tee{31} & 
\nodata		&
\nodata		\\

		&
\multicolumn{1}{r}{\rosat/HRI}	&
		&
 (0.1-2.4 keV)&	
		&
		\\


4		& 
1997 Feb 4-5 	& 
0.13\ppm0.02	& 
3\ud{3}{2}\tee{32} &
2.5\ppm0.5	&
0.3\ppm0.1	\\

		&
\multicolumn{1}{r}{\asca/GIS	}& 
		&
 (0.5-10  keV) 	&	
		&
		\\

5		 & 		
1999 Feb 9	&
\multicolumn{4}{c}{(see discussion of \S~\ref{sec:sax})}\\
		&
\multicolumn{1}{r}{SAX/LECS+MECS}	&
		&
 (0.5-10  keV) 	&	
		&
		\\

6		&
2000 June 23	&
0.176\ud{0.012}{0.015}	&
1.7\tee{32}	&
1.2\ud{0.4}{0.5}&
$<$0.06		\\
		&
\multicolumn{1}{r}{\chandra/ACIS-S} &
		&
 (0.5-10  keV) 	&	
		&
		\\
\enddata

\tablecomments{The quoted luminosity is the luminosity at the source
(i.e. -- unabsorbed, using the column density noted).  Assumed source
distance d=1.2 kpc; luminosities quoted from references which assumed
different distances have been changed.}

\tablerefs{
1,  \citenp{jvp87};
2, \citenp{asai96b}; 
3, \citenp{campana97};
4, \citenp{asai98}; 
5, \citenp{campana00}; 
6, present work. 
}
\end{deluxetable}

\begin{deluxetable}{ccccc|ccccc}
\scriptsize
\tablecaption{\label{tab:jointfit} H Atmosphere+Power Law Spectral Parameters of
\chandra/ACIS and \asca/SIS+GIS}
\tablewidth{0pc}

\tablehead{
\colhead{\nh} 		& 
\colhead{$\alpha$ }	& 
\colhead{\plfx $^b$}	&
\colhead{\ktH}		&
\colhead{$r$}		&
\colhead{Variable}	 &
\colhead{\chandra} 	&
\colhead{\asca} 	&
\colhead{\chisqrnu/dof} 	&
\colhead{prob.} 		\\
\colhead{(\ee{22} \percm)} 	& 
\colhead{}			& 
\colhead{}			&
\colhead{(keV)}			&
\colhead{(km) $(\frac{d}{1.2 {\rm kpc}})$}	&
\colhead{Parameter}		& 
\colhead{Value}			& 
\colhead{Value}			& 
\colhead{}			& 
\colhead{} 		\\
}
\startdata
(0.055)	&
1.0\ppm0.4		&
5.7\ud{1.0}{2.4}	&
\nodata			& 
12\ud{3.0}{2.4}		&
\ktH\			&
0.077\ppm0.007		&
0.085\ud{0.008}{0.007}	&
1.5/47			&
0.01			\\

(0.055)			&
1.7\ud{0.4}{0.2}	&
\nodata			& 
0.076\ud{0.005}{0.010}	&
12\ud{5}{1.3}		&
\plfx\			&
3.5\ud{1.6}{0.9}		&
9.1\ud{3.1}{2.1}			&
1.5/47			&
0.01			\\
\enddata
\tablecomments{$^b$Power-law fluxes are
unabsorbed (0.5-10.0 keV), in units of \ee{-13} \cgsflux; Values in
parenthesis are held fixed.}
\end{deluxetable}

\begin{deluxetable}{ccccc|ccccc}
\scriptsize

\tablecaption{\label{tab:3fit} H Atmosphere+Power Law Spectral Parameters of
\chandra/ACIS, \sax/LECS+MECS  and \asca/SIS+GIS}
\tablewidth{0pc}
\tablehead{
\colhead{} 		& 
\colhead{}		&
\colhead{}		&
\colhead{\chandra\ + \sax}	& 
\colhead{}	&
\colhead{}	&
\colhead{\asca}	 &
\colhead{} 	&
\colhead{} 	&
\colhead{} 		\\  \cline{4-5} \cline{7-8}
\colhead{\nh} 		& 
\colhead{\ktH}		&
\colhead{$r$}		&
\colhead{$\alpha$ }	& 
\colhead{\plfx $^b$}	&
\colhead{}	 &
\colhead{$\alpha$ }	& 
\colhead{\plfx $^b$}	&
\colhead{\chisqrnu/dof} 	&
\colhead{prob.} 		\\
\colhead{(\ee{22} \percm)} 	& 
\colhead{(keV)}			&
\colhead{(km) $(\frac{d}{1.2 {\rm kpc}})$}	&
\colhead{}			& 
\colhead{}			&
\colhead{}			& 
\colhead{}			&
\colhead{}			&
\colhead{}			& 
\colhead{} 		\\
}
\startdata
(0.055)	&
0.076\ppm0.007 	&
12.9\ppm2.6	&
1.0\ppm0.4	&
4.4\ud{1.9}{2.3}&
		&
1.9\ppm0.3	&
8.6\ppm1.6		&
1.0/52		&
0.43		\\
\enddata
\tablecomments{$^b$Power-law fluxes are
unabsorbed (0.5-10.0 keV), in units of \ee{-13} \cgsflux; Values in
parenthesis are held fixed.}
\end{deluxetable}

\end{document}